\documentclass[12pt]{iopart}
\usepackage{iopams} 
\usepackage {mathrsfs}
\usepackage {latexsym}
\usepackage {graphicx}
\usepackage {dsfont}
\usepackage {txfonts}
\usepackage {rotating}
\usepackage {wasysym}
\usepackage {multirow}
\usepackage {hhline}
\usepackage {hyperref}
\usepackage {color}
\usepackage {bm}
\usepackage{appendix}
\usepackage{acronym}
\usepackage{dcolumn}   
\usepackage {url}

\newcommand{\dcc}{LIGO-P1200062-v2}

\def\commitID{commitID: 706b261f620371aae9d4adc1ccc34bc932903642}
\def\commitDATE{Fri Sep 13 12:27:37 2013 +0100}

\newcommand{\trial}{trial} 
\newcommand{\trials}{\trial{s}} 
\newcommand{\trigger}{trigger} 
\newcommand{\triggers}{\trigger{s}} 

\newcommand{\signal}{\mathcal{H}^{+}}
\newcommand{\nosig}{\mathcal{H}^{-}}

\newcommand{\globpar}{\bm{\gamma}}

\newcommand{\nodet}{D^{-}}
\newcommand{\detect}{D^{+}}
\newcommand{\model}{\mathcal{H}}
\newcommand{\modelset}{\{\mathcal{H}\}}
\newcommand{\sigpar}{\bm{\theta}_{\model}}

\newcommand{\xth}{x_\mathrm{th}}
\newcommand{\dL}{d_{\mathrm{L}}}

\newcommand{\mum}{\mu_{\mathrm{m}}}
\newcommand{\sigm}{\sigma_{\mathrm{m}}}
\newcommand{\rhoth}{\rho_{\mathrm{th}}}

\begin{document}

\title{Avoiding selection bias in gravitational wave astronomy}
\author{C.~Messenger}
\ead{chris.messenger@astro.cf.ac.uk}
\address{School of Physics and Astronomy, Cardiff University, Queen's Buildings, The Parade, CF24 3AA}

\author{J.~Veitch}
\ead{johnv@nikhef.nl}
\address{Nikhef, Science Park 105, Amsterdam 1098 XG, Netherlands}
\date{\today}
\date{\commitDATE\\\mbox{\small \commitID}\\\mbox{\dcc}}

%
%
\begin{abstract}
  When searching for gravitational waves in the data from ground-based
  gravitational wave detectors it is common to use a detection
  threshold to reduce the number of background events which are
  unlikely to be the signals of interest.  However, imposing such a
  threshold will also discard some real signals with low amplitude,
  which can potentially bias any inferences drawn from the population
  of detected signals.  We show how this selection bias is naturally
  avoided by using the full information from the search, considering
  both the selected data and our ignorance of the data that are thrown
  away, and considering all relevant signal and noise models.  This
  approach produces unbiased estimates of parameters even in the
  presence of false alarms and incomplete data.  This can be seen as
  an extension of previous methods into the high false rate regime
  where we are able to show that the quality of parameter inference
  can be optimised by lowering thresholds and increasing the false
  alarm rate.
\end{abstract}

\maketitle


\acrodef{SNIa}[SNIa]{Type Ia supernovae}
\acrodef{pdf}[pdf]{probability density function}
\acrodef{GW}[GW]{gravitational wave}
\acrodef{SNR}[SNR]{signal-to-noise-ratio}
\acrodef{NS}[NS]{neutron-star}
\acrodef{BNS}[BNS]{binary neutron-star}


\section{Introduction}
%
%
Future generations of ground-based gravitational wave detectors, such
as the under-construction Advanced LIGO \cite{AdLIGO}, Advanced Virgo
\cite{AdVirgo} and Kagra \cite{Kuroda:LCGT} or the proposed Einstein
Telescope \cite{ET} are expected to detect a multitude of
gravitational wave sources in the coming years \cite{RatesPaper}.  The
analysis of these first detected signals will help to answer questions
about the rate of binary coalescence~\cite{Biswas:2009} and the
astrophysical distribution of neutron star and black hole
masses~\cite{mandel:multipleevents}.  As the number of observed
sources increases, it will be possible to use populations of
gravitational wave sources to perform cosmological parameter
estimation~\cite{delpozzo:cosmo,TaylorGairMandel:Cosmology,MessengerRead:2011,TaylorGair:2012}
and to test General
Relativity~\cite{delpozzo:massivegraviton,li:testinggr1,li:testinggr2}.
Previous searches for gravitational waves have used a hierarchical
pipeline based on a detection statistic, with a threshold applied to
this quantity to reduce the amount of data to be processed by later
stages of the pipeline~\cite{ihopepaper,InspiralS4}.  Because of the
very low amplitude of the expected signals relative to the detector
noise level, this process eliminates signals which fail to reach the
threshold as well as background noise, amounting to a selection of the
loudest events from all the true signals.  Any analysis which attempts
to draw inferences from a population of signals or events which have
been chosen in this way is vulnerable to \emph{selection bias} if the
population of detected signals does not match that of the underlying
sources. We draw a careful distinction between selection bias and
biases induced by an incorrect application of prior information
~\cite{HendrySimmons1990, LoredoHendry2010}, as the terms ``Malmquist
bias''~\cite{1913Eddington, 1920Malmquist, 1940Eddington} or
``Lutz-Kelker bias''~\cite{LutzKelker1973} are sometimes used to
describe both types of bias.

Many areas of astrophysics are subject to such effects, either by
intentional selection or thresholding to reduce the background while
performing data analysis, or by limited sensitivity of the instrument
(e.g. a flux limited survey~\cite{Loredo2007}).  Examples from the
literature include searches for \ac{SNIa}~\cite{Hicken2009},
\acp{GW}~\cite{InspiralS4}, extra-solar planets~\cite{UdrySantos2007}
and high energy neutrinos~\cite{Ageron2011,Abbasi2010}; and galaxy
cluster surveys~\cite{Allen2011,Cunha2009}.  Often the aim is to use
the resulting set of observations to infer global parameters of
interest, e.g. quantities that characterize the population
distribution. In the case of \ac{SNIa} for example, one may aim to
infer cosmological parameters from luminosity distance and redshift
measurements.  In cosmology, the traditional approach has been to
produce unbiased \emph{estimators} of individual event
parameters~\cite{Landy1992, Mantz2010, HendrySimmons1994,
  WillickStrauss1998, Lynden-Bell1988} which are then used as input
for inference of global quantities.  Thresholding has been used to
reduce the number of false alarms from a background population to a
tolerable level, for example thresholding on radio pulse width is
applied to data within the LOFAR NuMoon project ~\cite{Mevius2012}.
\cite{Martinez2011} applies a threshold on a fitting parameter to
optimize the separation of cosmic neutrinos and background for the
ANTARES neutrino experiment.  In \ac{GW}
searches~\cite{2010PhRvD..81j2001A}, thresholds are set on the
\ac{SNR} of an event to reduce the volume of data to be processed
before additional cuts are applied to minimize the false alarm rate
due to detector noise (which is another example of
selection). Candidates from future \ac{GW} searches will be used as
input for inferences on global parameters, such as the parameters of
the \ac{BNS} mass distribution.

It has been previously shown~\cite{Kelly2007}, that to avoid bias in
the analysis of astronomical data in a signal-dominated environment
(i.e. no false alarms), one should account for both the detections
\emph{and} the false dismissals within the experiment. In this paper
we provide a prescription for completely avoiding selection bias by
extending that approach beyond the signal dominated regime. To do this
we must consider the cases of detection and dismissal for all other
relevant models, most importantly the noise model. Our approach
accounts for the data which are thrown away by the application of the
threshold, making no assumption that they contain only noise, nor do
we assume that detected events are produced only by true signals.
Without this extension, future global inferences from GW signals would
be limited to the high-signal regime of \cite{Kelly2007}, which
discards all events that do not meet the threshold, resulting in
sub-optimal performance.  Our method will therefore be useful in
extracting the maximum information from a thresholded search, and is
also applicable to many other sensitivity-limited analyses. We
demonstrate this with an example, where lowering the threshold
increases the precision of parameter estimation for the \ac{BNS}
distribution and rate, which remains unbiased despite the presence of
false alarms.

%
%


%
%
\section{General Formulation}
%
%
Consider an experiment in which many measurements are made,
each producing data which may contain potential detections,
and which are then compared with a detection threshold. We refer to any
measurements which pass the threshold as ``\triggers'',
since in most practical cases we will
include one or more noise-only models alongside the true signal model.
Any measurements which do not pass the threshold are discarded.
These non-\trigger{} cases also carry information about
the source population, and one should not assume the lack of a \trigger{}
implies the lack of true signal, or vice-versa.

We divide the experiment into statistically independent multiple measurements, and denote the outcome of each as $D$.
In practice this might mean dividing a search into multiple time periods.
The threshold is set on the measured quantity $x$, which we will call the ``detection statistic''.
In general $x$ may be a maximum-likelihood statistic or simply a direct measurement of a
physical parameter value but in all circumstances 
will be subject to noise.

If the statistic passes a threshold $x>\xth$ a \trigger{} is
produced, and the value of $x$ is recorded; we denote this case
$\detect$, and $D=(\detect,x)$.  If on the other hand
$x\le\xth$, no
\trigger{} is produced and no value for $x$ is recorded and
$D=\nodet$.  We define $\modelset$ as our set of plausible models and
must include at least two models with one describing the presence and
the other the absence of a signal.  In general the method extends to
many alternative signal models. When specifying a set of models in this way
we allow that any model may be true, and that the \trigger{} has a non-zero
probability under all of the models. For any single model $\model$ the likelihood function
defining the distribution of the statistic $x$ is expressed as
$p(x|\model,\sigpar,\globpar,I)$ where $\sigpar$ are the parameters
unique to this model, and $\globpar$ are the \emph{global} parameters
which are common to all \trials{} (following standard Bayesian
notation we use $I$ throughout to represent additional information regarding the experiment, and the global parameters
 $\globpar$).
Upon performing the single \trial{}, we are
interested in inferring the parameters $\globpar$.  In the case of a
statistic above threshold, $\detect$, we can write the posterior
distribution on $\globpar$ as
\begin{eqnarray}
  p(\globpar|\detect,x,I)&=&\sum_{\modelset}p(\globpar,\model|\detect,x,I)\nonumber\\
  &=&\frac{p(\globpar|I)}{P(x|I)}\overbrace{\sum_{\modelset}p(x|\globpar,\model,I)P(\model|\globpar,I)}^{\mathrm{\trigger{}\,
      likelihood}\,p(\detect,x|\globpar,I)}\label{e:pos_det}
\end{eqnarray}
where $P(\model|\globpar,I)$ represents the prior probability on the
model $\model$ conditional on the global parameters.
Here we are summing the trigger likelihoods over the set of possible models $\modelset$, and
we do not assume that the model priors are independent of the global parameters,
which will be important when discussing signal
abundance and rates. The range of possible models can include those where more than one
signal is present in the data. We have used the fact that the
probability of a \trigger{} conditional on the value $x$ is strictly
equal to $H(x-\xth)$ (where $H(x)$ is the Heaviside step function)
which is equal to unity for $x>\xth$.
 If the model has local parameters $\sigpar$, we calculate the
factor $p(x|\globpar,\model,I)$, i.e. the likelihood of the data
conditional on the model and the global parameters, by
marginalization:
\begin{equation}
  p(x|\globpar,\model,I)=\int\limits_{\Theta_{\model}} d\sigpar
  p(x|\sigpar,\globpar,\model,I)p(\sigpar|\globpar,\model,I),
\end{equation}
where $\Theta_{\model}$ is the local parameter space associated with
the local parameter $\theta_{\model}$ and the model $\model$.
Note that we do not modify this likelihood function to reflect our
knowledge of the threshold.

In the alternative case $\nodet$ where we discarded the 
specific value of $x$, we do still know that $x\le\xth$. For a
non-\trigger{} we can therefore write the posterior probability on
$\globpar$ (analogous to Eq.~(\ref{e:pos_det})) as
\begin{equation}
  p(\globpar|\nodet,I)=\frac{p(\globpar|I)}{P(\nodet|I)}\overbrace{\sum_{\modelset}P(\nodet|\globpar,\model,I)P(\model|\globpar,I)}^{\mathrm{non-\trigger{}\,likelihood}\,p(\nodet|\globpar,I)}\label{e:pos_nodet}
\end{equation}
As the value of $x$ is unknown the posterior on
$\globpar$ must be marginalized over both the local parameters
$\sigpar$ and the unknown data $x$ in the range $x\le\xth$. Therefore
only the information $\nodet$ appears in the likelihood of a
non-\trigger{} conditional on the model and the global parameters:
\begin{eqnarray}
  P(\nodet|\globpar,\model,I)&=&\int\limits_{\Theta_{\model}}d\sigpar\hspace{-0.1cm}\int\limits_{x{\le}\xth}
  \hspace{-0.1cm}dx~p(x|\sigpar,\globpar,\model,I)p(\sigpar|\globpar,\model,I)\nonumber \\\label{eq:pnodet}
\end{eqnarray}
The quantity derived in
Eq.~(\ref{eq:pnodet}) is related to what is known as the
\emph{selection function} or \emph{detection efficiency}, defined as
the probability of a trigger given the true signal parameters for a
given signal model.  Here it appears in a modified form: the probability of a
\emph{non-\trigger{}} given a set of global parameters and a model.
It is only in this term that the detection threshold $\xth$ actually
appears.
In practice this
quantity (just like the detection efficiency) can be difficult to
estimate in some cases.

%
%
\subsection{An ensemble of \trials{}}
Using Eqns.~(\ref{e:pos_det})
and~(\ref{e:pos_nodet}) together with Bayes' theorem we are able to
isolate the single \trial{} likelihood functions (identified by the
over-braces) in both the \trigger{} and non-\trigger{} cases,
$P(\detect,x|\globpar,I)$ and $p(\nodet|\globpar,I)$ respectively.  We
can now also consider analysing an ensemble of $N$ \trials{} denoted
as $\{D\}$, where we index each $D_{i}$ using $i\in(1,N)$. The
posterior probability for the global parameters after $N$ independent
\trials{} are performed is simply
%
\begin{equation}
p(\globpar|\{D\},I)=\frac{p(\globpar|I)}{P(\{D\}|I)}\prod\limits_{i=1}^{N}p(D_i|\globpar,I).\label{e:pos_multiple1}
\end{equation}
We can further decompose this expression into \trigger{} $\detect$ and
non-\trigger{} $\nodet$ factors,
\begin{equation}
p(\globpar|\{D\},I)=\frac{p(\globpar|I)}{P(\{D\}|I)}\prod_{j=1}^{n}p(\detect,x_{j}|\globpar,I_{j})\prod\limits_{k=1}^{N-n}P(\nodet|\globpar,I_{k}),\label{e:pos_multiple2}
\end{equation}
where $j$ and $k$ index the $n$ \triggers{} and the $N-n$
non-\triggers{} respectively, and $I_j$, $I_k$ allow for different
background information (such as time-varying detector sensitivity) in each measurement.

%
%
We now focus on the common case where there are only two possible
models: the presence ($\signal$) or absence ($\nosig$) of a single
signal per \trial{}. We can reach this case in practice by considering a trial
to be a period of time where the probability of $>1$ signal is arbitrarily
small.
Equation~(\ref{e:pos_multiple2}) becomes
\begin{eqnarray}
  p(&\globpar&|\{D\},I)=\frac{p(\globpar|I)}{P(\{D\}|I)}\nonumber\\
 &&\times\prod\limits_{j=1}^{n}\left[p(\detect,x_{j}|\globpar,\signal,I_{j})P(\signal|\globpar,I_{j})+p(\detect,x_{j}|\globpar,\nosig,I_{j})P(\nosig|\globpar,I_{j})\right]\nonumber\\
 &&\times\prod\limits_{k=1}^{N-n}\left[P(\nodet|\globpar,\signal,I_{k})P(\signal|\globpar,I_{k})+P(\nodet|\globpar,\nosig,I_{k})P(\nosig|\globpar,I_{k})\right].\label{e:master}
\end{eqnarray}
To relate this to more standard statistical terms, consider the key
components which make up the expression, each of which are in
reference to single trials:
\begin{itemize}
\item $p(\detect,x_{j}|\globpar,\signal,I_{j})$ is the likelihood of
  producing a \trigger{} with statistic value $x_{j}$ when a signal is
  truly present.
\item $p(\detect,x_{j}|\globpar,\nosig,I_{j})$ is the likelihood of
  producing a \trigger{} with statistic value $x_{j}$ from only the
  background distribution.
\item $P(\nodet|\globpar,\signal,I_{k})$ is the likelihood that a
  true signal does not produce a
  \trigger{}, i.e. the \emph{false dismissal likelihood} (or
  $1-$detection efficiency).
\item $P(\nodet|\globpar,\nosig,I_{k})$ is the likelihood that
  no \trigger{} is produced when no signal is present,
  i.e. the \emph{true dismissal likelihood}.
\end{itemize}

%
%
%
The abundance or true rate of signal events is important in the two
model case, when the two models are for a single signal and
background.  As the two models give a complete description of the
data, we can write $P(\signal|\globpar,I)=1-P(\nosig|\globpar,I)$.
The actual rate of signals, or a quantity from which it can be
derived, must be included as one of the global parameters $\globpar$,
and if it is not known then it can be estimated from the ensemble of
\trials{}.  This makes the prior probabilities for the models
dependent on the global parameters $\globpar$, so the joint posterior
distribution for the rate will not in general be separable from the
rest of global parameters.  Alternatively, if the true rate of signals
is known, it can be substituted into these expressions instead.
However, using incorrect assumptions about the rate parameter will
therefore bias inference on other global parameters.

%
%
The key idea of our approach (in terms best suited to the two model
problem) is to use the true and false dismissal probabilities to
incorporate the information from the absence of \triggers{}. This
requires that these quantities be either calculated or measured.
Without these, any observations which can produce non-\triggers{}
risk being either biased or sub-optimal and may require ad-hoc
corrections. It is only by acknowledging and accounting for ignorance
that we can derive the correct results.  Equation~(\ref{e:master})
contains all the parts necessary to infer global parameters without
bias, making full use of the information that is kept, and accounting
for that which is deliberately (or necessarily) discarded.


\section{A gravitational wave example}
\begin{figure}
  \begin{center}
    \includegraphics[width = \columnwidth]{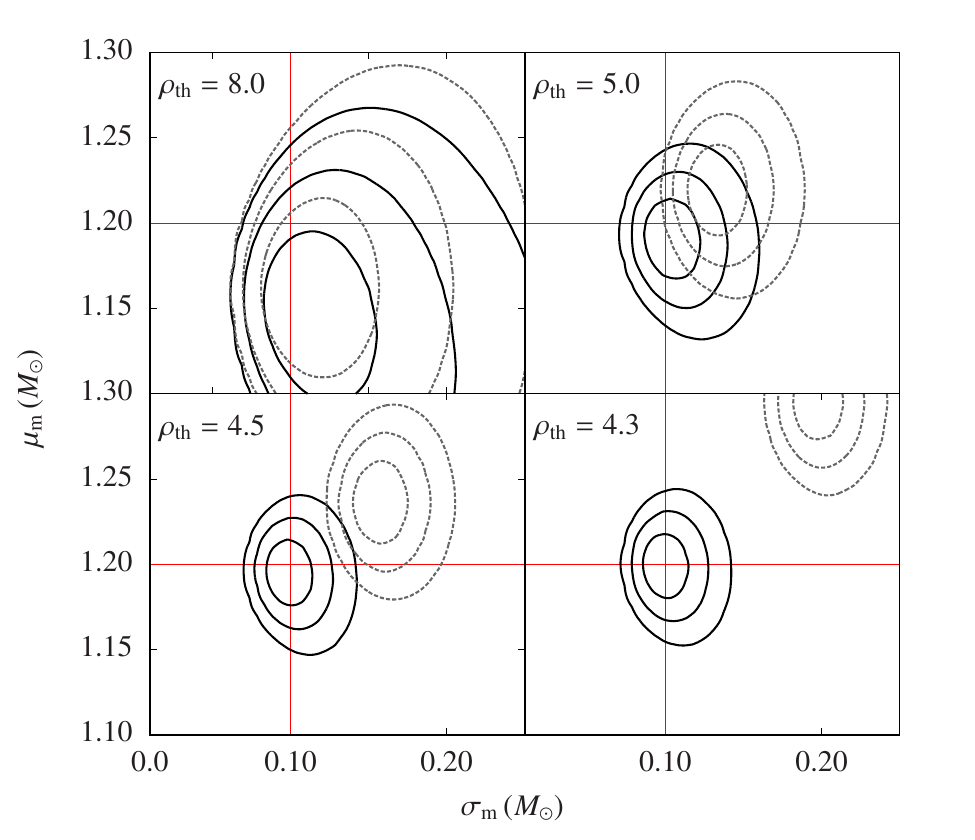}
    \caption{Joint probability contours on the combined global
      parameters $\mu_{\mathrm{m}}$ and $\sigma_{\mathrm{m}}$ for $4$
      different threshold values $\rhoth=8,5,4.5,4.3$. There were a
      total of 525600 experiments of which 87 contained signals.  For
      $\rhoth=8$ there were 6 \triggers{} and all were signals.  For
      $\rhoth=5$ there were 29 \triggers{} of which 25 were signals
      and 4 were noise events.  For $\rhoth=4.5$ there were 48
      \triggers{} of which 37 were signals and 11 were noise events,
      and finally, for $\rhoth=4.3$ there were 78 \triggers{} of which
      40 were signals and 38 were noise events.  The true values of
      the mean and standard deviation of the mass are indicated by the
      solid red lines. Contours enclose 68\%, 95\% and 99.7\% of the
      probability.  The black contours correspond to the application
      of our method accounting for both \triggers{} and
      non-\triggers{} and both the signal and noise models.  The
      dashed grey contours represent results from an analysis
      accounting for \triggers{}-only and assuming a signal model
      only.\label{fig:GWexample2D}}
  \end{center}
\end{figure}
\begin{figure}
  \begin{center}
    \includegraphics[width = \columnwidth]{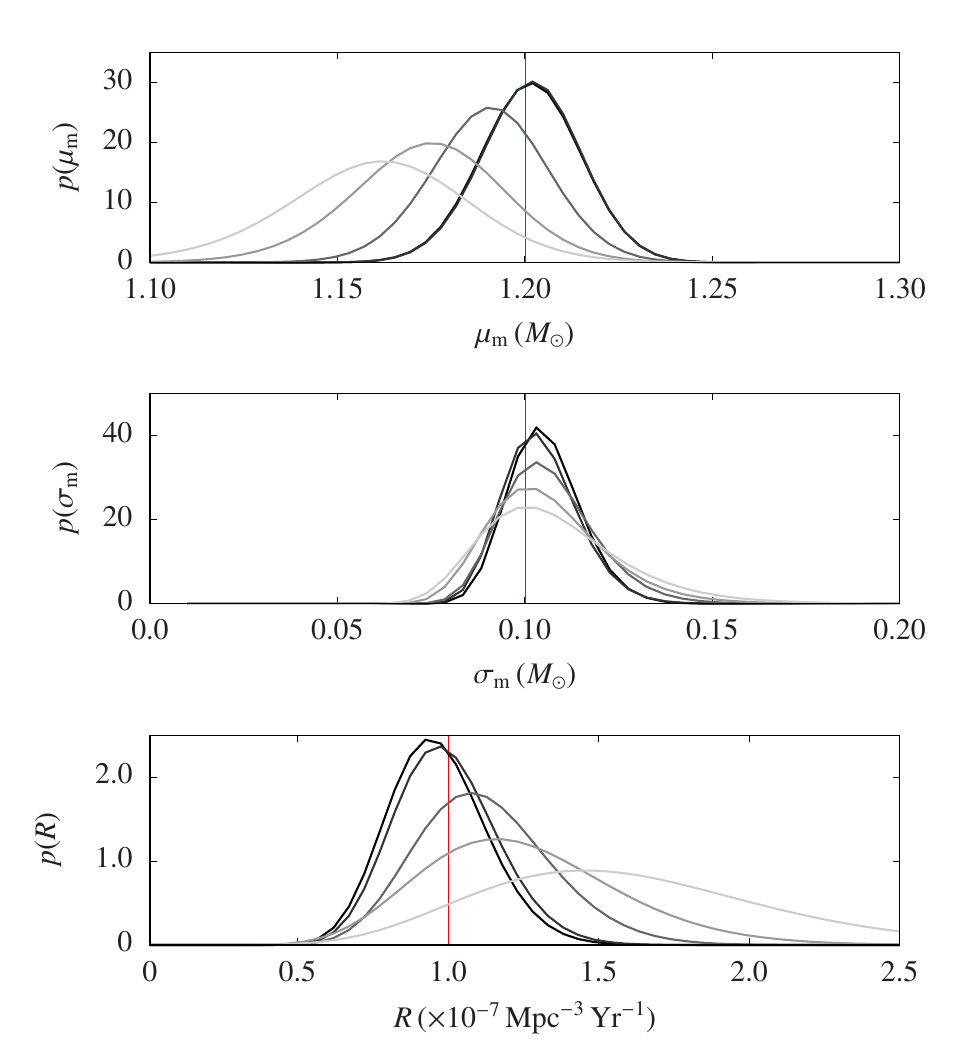}
    \caption{Marginalised posterior probability distributions on the
      global mass parameters $\mu_{\mathrm{m}}$, $\sigma_{\mathrm{m}}$ and
      the rate $R$ for a range of different threshold values
      $\rhoth=7,6,5,4,3$.  The fraction of noise \triggers{}
      above threshold associated with
      these values are $0,0,0.138,0.80,0.988$ respectively.
      There were a total of 525600 experiments of which 87 contained
      signals and the curves go from light grey to black as the
      threshold decreases. The true values of the global parameters
      are indicated by the solid vertical red
      lines. \label{fig:GWexample1D}}
  \end{center}
\end{figure}
%
%
Searches for \acp{GW} signals from compact binary coalescences employ
a threshold when selecting which triggers to keep, and so
astrophysical statements drawn from these observations will be subject
to selection effects~\cite{2010PhRvD..81j2001A}.  To show how our
method can be used to avoid bias, we will consider a toy model in
which a \ac{GW} detector searches for \acp{BNS}, and we want to
infer the global parameters: the event rate per unit co-moving volume
$R$, and the mean $\mum$ and standard deviation $\sigm$ of the
component mass distribution.
We assume that the population of component masses is governed by a
Gaussian distribution, and that a single trial constitutes an all-sky
search over the duration $\Delta t=60$ seconds. 
%
%
For each \trial{}, the data comprises a measurement of the \ac{SNR}
$\rho$ and estimates of the component masses $\mathbf{m}=(m_{1},m_{2})$ of any
event that exceeds the \ac{SNR} threshold $\rho_{\mathrm{th}}$.  In
reality, the \ac{SNR} (our detection statistic) is generated through
matched filtering of the data.  This process includes both analytical
maximization of the \ac{SNR} over one or more of the local nuisance
parameters e.g. phase at coalescence, and numerical maximization over
a bank of templates on the component mass space.  For simplicity, we
assume the component mass estimates to be noise free exact
measurements although in practice there will be some uncertainty in
their values.  In addition we ignore the complications associated with
the template bank search over the component mass space noting that in
practice, maximization of a detection statistic over a bank of
templates would act to modify the distribution of the detection
statistic for both the signal and noise models.  Although our model is
simple and lacks the detail of real observations, it will serve to
show the salient points of the method.

%
%
We suppose that events arise through either a signal model $\signal$
or through a noise model $\nosig$.  According to the signal model we
define the expected (optimal) \ac{SNR}-squared of a source as
\begin{equation}\label{eq:snrsq}
  \rho_{\mathrm{opt}}^{2} = \frac{5\pi^{-4/3}}{96}\frac{\mathcal{M}^{5/3}}{\dL^{2}}\int_{20}^{1500}\frac{f^{-7/3}}{S_{h}(f)}df
\end{equation}
where the chirp-mass $\mathcal{M}=M\eta^{3/5}$ with the total mass
$M=m_{1}+m_{2}$ and the symmetric mass ratio $\eta=m_{1}m_{2}/M^{2}$.
The luminosity distance is given by $\dL$ and the detector noise
spectral density is denoted by $S_{h}(f)$ for which we use the
anticipated Advanced LIGO noise
curve~\cite{2002grg..conf...72C}.  Under the signal model, we assume
that the measured \ac{SNR}-squared is drawn from a non-central $\chi^{2}$
distribution with 2 degrees of freedom and a non-centrality parameter
$\rho_{\mathrm{opt}}^{2}$.  The component masses are generated from the
prior distributions governed by the global parameters $\mum,\sigm$ and
are measured without noise.
Under the noise model, we assume the measured \ac{SNR}-squared is drawn
from a central $\chi^2$ distribution with 2 degrees of freedom and
should be interpreted as arising from detector noise.  All data
recorded above the \ac{SNR} threshold due to the noise model will have
corresponding component mass measurements drawn from a uniform
distribution of masses within our prior search range defined as
$\mathcal{S}_{m}\in(0.9M_{\odot},1.8M_{\odot})$.  Hence, a noise model
trigger will be associated with any possible component mass
configuration with equal probability, unlike the signal model events
that follow the astrophysical prior distribution.
In practice the distribution of inferred masses caused by the noise
background will follow its own distribution which can be modelled or
measured.

%

We assume a uniform distribution of sources in volume giving a prior
distribution on the luminosity distance of
$p(\dL|\globpar,I)=3\dL^2/(\dL^{\mathrm{max}})^{3}$. We have
used $\dL^{\mathrm{max}}=600$\,Mpc, chosen to be beyond the distance at
which BNS systems will be detectable, and ignored cosmological
effects since the reach of Advanced LIGO is $z{\sim}0.05$ for
\acp{BNS}.
%
We simulate $1$ year of \trials{} with a true astrophysical event rate
(per unit volume per unit observation time) of
$R=10^{-7}\,\mathrm{Mpc}^{-3}\mathrm{yr}^{-1}$ with component mass
distribution parameters $\mum=1.2\,M_{\odot}$ and
$\sigm=0.1\,M_{\odot}$. 

%
%
First consider the probability of obtaining a non-\trigger{} per trial
under the assumption of the signal model $\signal$,
\begin{eqnarray}
  P(\nodet|\globpar,&&\signal,I)=\int\limits_{0}^{\rhoth^2}\hspace{-0.1cm}dx\hspace{-0.1cm}\int\limits_{\mathcal{S}^{2}_{m}}\hspace{-0.1cm}d\mathbf{m}'~p(x,\mathbf{m}'|\globpar,\signal,I)\nonumber\\
&=&\int\limits_{0}^{\rhoth^{2}}\hspace{-0.1cm}dx\,\frac{1}{2}\exp\left(-\frac{1}{2}\left[x+\rho_{\mathrm{opt}}^{2}(\mathbf{m})\right]\right)I_{0}\left(x\rho_{\mathrm{opt}}(\mathbf{m})\right) \label{eq:exnodetsig}
\end{eqnarray}
where $I_{0}$ is a modified Bessel function of the first kind and
$x,\mathbf{m}'$ represents the unknown true squared \ac{SNR} and
unknown true component
masses over which we marginalize.  The complementary quantity, the
probability density associated with detection for the signal model,
for the $j$'th \trial{} is
\begin{eqnarray}
p(\rho^{2}_{j},&&\mathbf{m}_{j}|\globpar,\signal,I)=\int\limits_{\mathcal{S}^{2}_{m}}\hspace{-0.1cm}d\mathbf{m}'~p(\rho^{2}_{j},\mathbf{m}_{j},\mathbf{m}'|\globpar,\signal,I)\nonumber\\
&=&\frac{1}{2}\exp\left\{-\frac{1}{2}\left(\rho^{2}_{j}+\rho_{\mathrm{opt}}^{2}(\mathbf{m}_{j})\right)\right\}I_{0}\left(x\rho_{\mathrm{opt}}(\mathbf{m}_{j})\right)\label{eq:exdetsig}
\end{eqnarray}
where we have used $p(\mathbf{m}|\mathbf{m}',\globpar,\signal,I) =
\delta(m_{1}-m'_{1})\delta(m_{2}-m'_{2})$ since the mass measurements
are noise free in this example.  For non-\trigger{} and \trigger{}
cases under the noise model we have
\begin{eqnarray}
  P(\nodet|\globpar,\nosig,I)&=&\int\limits_{0}^{\rhoth^2}\hspace{-0.1cm}dx\hspace{-0.1cm}\int\limits_{\mathcal{S}^{2}_{m}}\hspace{-0.1cm}d\mathbf{m}'~p(x,\mathbf{m}'|\globpar,\nosig,I)\nonumber\\
&=&1 - \exp\left(-\rhoth^{2}/2\right) \label{eq:exnodetnoise}
\end{eqnarray}
\begin{eqnarray}
 p(\rho^{2}_{j},\mathbf{m}_{j}|\globpar,\nosig,I)&=&\int\limits_{\mathcal{S}^{2}_{m}}\hspace{-0.1cm}d\mathbf{m}'~p(\rho^{2}_{j},\mathbf{m}_{j},\mathbf{m}'|\globpar,\nosig,I)\nonumber\\
&=&\frac{1}{2\Delta m^2}\exp\left(-\rhoth^{2}/2\right)\label{eq:exdetnoise}
\end{eqnarray}
where $\Delta m=0.9M_{\odot}$ is the prior range of component masses
and we have used the noise model component mass prior distribution
$p(\mathbf{m}'|\globpar,\nosig,I) = 1/\Delta m^{2}$.  This is also
equal to the noise model component mass likelihood
$p(\mathbf{m}|\mathbf{m}',\globpar,\nosig,I)$.
%
%
%
We define the model priors assuming that $\Delta t$ is sufficiently
small that within a volume $V$, $P(\signal|\globpar,I)=RV\Delta{}t \ll
1$, so $P(\nosig|\globpar,I)=1-RV\Delta{}t$.

In Fig.~\ref{fig:GWexample2D} we show the joint posterior probability on
the global parameters $\mum$ and $\sigm$ for a set of different
\ac{SNR} threshold values.  The model can be made more complicated by including
any number of extra parameters, global or local, e.g. cosmological parameters or the nuisance
parameters of individual sources.
As the threshold is decreased we increase the number of
\triggers{} due to signal events but also increase the number of false
alarms (\triggers{} due to noise events).  For the case where
$\rhoth=8$ the number of false alarms is zero but for $\rhoth=4.3$
the number of false alarms is nearly equal to the number of true
alarms~(see figure caption for details).
The solid contours in Fig.~\ref{fig:GWexample2D} show the
results of applying our method to the data where it can be seen that
decreasing the threshold and correctly accounting for the increase in
true and false alarms improves parameter estimation accuracy whilst
remaining consistent with the correct parameter values.

An additional result is included in dashed grey lines showing the
biases that can occur when failing to incorporate all relevant information
into an analysis.
This incorrect result is calculated by evaluating
Eq.~\ref{eq:exdetsig} for each \trigger{} and therefore ignores the
information contained within both the model prior and the
non-\triggers{}.  Although it performs as expected when the threshold
is high enough to allow no false alarms, the subsequent global
parameter estimates become strongly biased for both mass parameters as
the threshold decreases and the false alarm rate increases. This
analysis assumes that the wider distribution of masses caused by the
noise model reflects the real signals, leading to an overestimate of
the mass standard deviation. The mean mass is also overestimated since
the average mass measurement due to the noise model is equal to the
centre of the mass prior range which in this case is higher than the
true value. The specific behaviour depends on the nature of the
signal, background, and the (incorrect) assumptions that are made, but
this is illustrative of the approach implicitly taken by those studies
which assume that a given set of gravitational wave \triggers{} will all
be due to the signal model.

We also show the precision of the global parameter estimates as a
function of the threshold value.  This is seen in
Fig.~\ref{fig:GWexample1D} where the marginalised posterior
distributions on the global parameters including the rate $R$ are
plotted for a range of threshold values. We see that the posterior
distributions become narrower as the threshold is decreased where more
signals \emph{and} noise events are recorded above threshold.  We
stress that this improvement in the precision of global parameter
estimation is unbiased by the presence of false alarms.  This is only
true when \triggers{}, non-\triggers{}, signal model, and the noise
model are correctly incorporated in the construction of the global
parameter posteriors. It is also notable that the reduction in
posterior width does not continue linearly with the reduction in
threshold or with the corresponding increase in false alarms.  In
fact, further decreasing the threshold yields diminishing
improvements, and the resulting distribution asymptotically approaches
the optimal one which would be obtained by removing the threshold
altogether.  In that case the presence of noise still limits the
precision attainable even though all the data is analysed and none
discarded.  This indicates that there exists a sensible choice of
threshold value at which additional computation expense in processing
\triggers{} from lower thresholds returns little information.  It also
implies that operating at thresholds corresponding to high false alarm
values will yield the most precise parameter estimates.


\section{Discussion}
%
%
We have derived a general formalism for analysing data which have been
subjected to a threshold, which accounts for both the knowledge of and
the ignorance of the actual result of an experiment. This formalism
was applied to a simplified real-world example problem from the field
of \ac{GW} data analysis which will be encountered when sufficient
detections have been made.
The method succeeds for any number of false alarms and reduces
to known formulae in limiting cases, such as zero
false alarm rate~\cite{Kelly2007}.  We find that knowledge of the true
rate or abundance of signals represents crucial information when
making inferences with an ensemble of \trials{}, and that it can
therefore be inferred from them.
Prior knowledge of signal and noise models must be incorporated, but measured values of
efficiency and false alarm rates can be used when they are difficult
to derive analytically.

Our work has implications for research which uses sets of
gravitational wave signals to probe astrophysical, cosmological, or
gravitational parameters, of the kind which have already been
suggested in the
literature~\cite{delpozzo:cosmo,TaylorGairMandel:Cosmology,MessengerRead:2011,TaylorGair:2012,delpozzo:massivegraviton,li:testinggr1,li:testinggr2}. We
have shown that the best global parameter estimates will come from
using the lowest threshold allowed by computational limitations, which
will include using events which were (probably) generated by the
background noise.

Our method could be applied in many other noise-limited fields of research,
ranging from particle physics to cosmology. 
%


\ack
The authors are grateful to 
J.~Clark, T.~Dent, S.~Fairhurst, I.~Harrison, M.~Hendry, D.~Keitel, T.~G.~F.~Li,
W.~Del Pozzo, G.~Pratten, R.~Prix, C.~R\"over, B.~S.~Sathyaprakash, P.~Sutton, M.~Vallisneri
C.~Van~Den Broeck, and G.~Woan for useful discussions and comments.

\section*{References}
\bibliographystyle{unsrt}
\bibliography{masterbib}

\end{document}